\documentstyle[12pt,epsfig]{article}

\begin{document}

\rightline{FNT/T-96/8}

\begin{center}
%\noindent
{\large\bf  $ {\cal O} ( \alpha^2  ) $ Next-to-Leading  Photonic 
Corrections  to  Small-Angle Bhabha 
Scattering   in the 
Structure Function Formalism$^*$}
\vskip 4pt
%{\large }
\end{center}

\noindent
\begin{center}
Guido MONTAGNA$^{a,b}$, Oreste NICROSINI$^{b,a}$  
and Fulvio PICCININI$^b$ 
\end{center}

\noindent
\begin{center}
$^a$ Dipartimento di Fisica Nucleare e Teorica, 
Universit\`a di Pavia, Italy \\
%\end{center}
%\begin{center}
$^b$ INFN, Sezione di Pavia,  Italy  
\end{center}

\begin{abstract}{\small A general method for computing 
$ {\cal O} ( \alpha^2 ) $ 
and higher-order next-to-leading 
photonic corrections is presented and applied 
to the precision 
calculation of the small-angle Bhabha scattering
cross section in the phase-space region of interest for the luminosity
measurement at LEP. The formulation 
is based on a proper matching of exact
$\cal O (\alpha)$ results with higher-order corrections in the Structure
Function formalism.  
The results of the approach are analytically compared with  
theoretical calculations, both for $s$- and
$t$-channel  processes, available for simple Event Selections. 
Numerical predictions  for realistic Event Selections are also 
provided and critically compared with the ones existing in the literature. }
\end{abstract}

%%%\vskip 24pt
%%%\begin{center}
%%%%%          Submitted to Physics Letters B
%%%\end{center}

\vskip 48pt \noindent
E-mail: \\
montagna@pavia.pv.infn.it \\
nicrosini@pavia.pv.infn.it, nicrosini@vxcern.cern.ch \\
piccinini@pavia.pv.infn.it \\

\vfil
\leftline{FNT/T-96/8}
\leftline{May 8, 1996}

\vfil \hrule \footnotesize \noindent
$^*$Based on a talk given at the third
meeting of the ``Event Generators for Bhabha Scattering'' Working Group,
Workshop on Physics at LEP2, CERN, 31.05.1995.
\eject
\normalsize
%%%%%%%%%%%%%%%%%%%%%%%%%%%%%%%%%%%%%%%%%%%%%%%%%%%%%%%%%%%%%%%%%%%%%%%%%
\leftline{\large\bf 1. Introduction}
\vskip 12pt
Small-angle Bhabha (SABH) Scattering is used at $e^+ e^- $ colliders to measure
the accelerator luminosity. Since the experimental error at present reached by
the LEP Collaborations is better than 0.1\%~\cite{experr}, in order to exploit 
such an experimental achievement it is mandatory to provide
theoretical predictions for the cross section in the luminosity region at the
same level of accuracy, or even better. 

During 1995, within the Workshop on 
``Physics at LEP2'' at CERN, the Working Group ``Event Generators for Bhabha
Scattering'' took place~\cite{bharep}. The major task of the Working Group was
to make an inventory of all the available Monte Carlo (MC) and non-MC codes for
the computation of the SABH cross section, with the aim of reaching a deeper
understanding of the SABH process. The main result of the Working Group was a
substantial reduction of the theoretical error on the 
SABH cross section~\cite{bharep,comdoc} 
from 0.16\% to 0.11\%, but further improvements are 
demanded. In particular, at the present stage the dominant part of the
theoretical error comes from the ${\cal O}  ( \alpha^2 L)$ photonic
corrections, where $L = \ln ( Q^2 / m^2 )$ is the usual collinear logarithm, 
which for the time being are not fully under  control for a 
realistic experimental Event Selection (ES). 

The main tool by means of which such a result has been achieved, was the
critical comparison of all the available theoretical formulations and the
corresponding numerical codes~\cite{bhagen, bhlumi, nllbha,  sabsyr,
sabscpc}, and the understanding of their
differences, where present. Given the relevance of this subject, we considered
important to improve our theoretical formulation presented in~\cite{sabsyr},
with the goal of providing a contribution to a further reduction of the
theoretical error. The aim of the present note is to describe the recent 
theoretical developments of our approach to the understanding of the SABH 
process, with particular emphasis on ${\cal O}  ( \alpha^2 L)$ and
higher-order next-to-leading photonic
corrections.  
The results of the approach are analytically compared with  
theoretical calculations, both for $s$- and
$t$-channel  processes, available for simple ES's. 
Numerical predictions  for realistic ES's are also 
provided and critically compared with the ones existing in the literature.

From now on, we will focus our attention on QED radiative corrections to the
dominant part of the cross section in the small-angle regime, namely  the cross
section for $t$-channel photon exchange. 
%A first approach to this problem has been
%done in~\cite{sabsyr}, under the assumption that a specific cut on the final
%state fermions is applied, namely a cut on the energies of the fermions.
%The present formulation allows one, 
%on the contrary, to impose arbitrary cuts on
%the final state fermions.
The corrections due to the other
non-dominant contributions, namely $s$-channel diagrams, $\gamma$-$Z$
interferences and so on, can be accounted for at the leading logarithmic (LL) 
level, as shown in~\cite{sabsyr}. 
We will start considering Bare ES's. They are
absolutely unrealistic from the point of view of luminometry at LEP, so their
interest is, in some sense, academical. In spite of this, they however
represent a useful benching situation, since the only available calculation
complete at the level of ${\cal O}  ( \alpha^2 L)$ corrections~\cite{nllbha} 
is conceived for such ES's. 
The results concerning Calorimetric ES's, the truly realistic ones,
will be derived as special cases of the ones obtained for Bare ES's: as a first
step, final-state radiation will be switched off in the LL part of the result,
consistently with the fact that,  for an inclusive measurement, final-state
mass ``singularities'' are absent; next, it will be taken into account by means
of a proper definition of the final-state radiation factor.  

%Moreover, only the formulation for 
%calorimetric ES's will be described in some detail here, 
%being the truly realistic one: 
%for such ES's,
%the final-state QED corrections are absent at the LL level, 
%leading  to a noticeable simplification in the theoretical formulation. Bare
%ES's, on the contrary, are technically more involved and are completely
%unrealistic from the point of view of luminometry at LEP. So their interest is
%essentially academical; nonetheless they will be considered shortly, when
%showing some numerical results, since they
%represent a useful benching situation. 
\vskip 12pt\noindent  
\leftline{\large\bf 2. Theoretical formulation}
\vskip 12pt
In order to settle down the basic notations, let us recall the form of the
corrected cross section in the LL approximation within the Structure Function
(SF)  formalism~\cite{sf}: 
\begin{eqnarray}
\sigma_{LL}   && = \int d x_1 d x_2 d y_1 d y_2 \int d I_{cm} 
D_{\beta_i}(x_1) D_{\beta_i}(x_2) \nonumber \\
&&{{d \sigma} \over {d I_{cm}}} ({I}_{cm} (x_1, x_2, y_1, y_2 )) 
D_{\beta_f}(y_1) D_{\beta_f}(y_2) . 
\label{eq:sfc}
\end{eqnarray}
In eq.~(\ref{eq:sfc}), $ D_{\beta_i}(x_{1,2})$ are the usual electron 
and positron structure 
functions, describing initial-state radiation, as can be found
in~\cite{sabsyr}; $ D_{\beta_f}(y_{1,2})$ are the analogous ones for
final-state radiation. In general, $\beta_i$ and $\beta_f$ are the radiation
factors for initial- and final-state radiation, respectively; in the case of a
Bare ES, one has 
$\beta_i = \beta_f = 2 (\alpha / \pi) \left[ \ln (-t / m^2 ) - 1
\right]$. The case of a Calorimetric ES will be discussed later on.   
$d \sigma / d I_{cm}$ is the density to be QED corrected, in the
centre of mass (cm) reference frame after initial state radiation. 
$I(x_1, x_2, y_1, y_2 )$
represents a convenient set of independent variables, able to describe the
configuration of the event corresponding to the kernel cross section, in
presence of initial- and final-state radiation.  
It is understood
that the event generated in the cm frame after initial state radiation is
boosted along the beam-axis according to the amount of radiation lost
longitudinally by the incoming particles, then modified by the effect of
final-state radiation, and then accepted or rejected
according to a given ES in the laboratory frame. 

The $\cal O (\alpha)$ content of eq.~(\ref{eq:sfc}) can be written as 
\begin{eqnarray}
&&\sigma_{LL}^{(\alpha)}   =  \int d x  \int d I_{cm} 
{{1 + x^2} \over {1-x}} \Bigg\{ {1 \over 4} 
\beta_i  \Bigg[
{{d \sigma} \over {d I_{cm}}} ({I}_{cm} (x, 1,1,1)) \nonumber \\
&&  +  
{{d \sigma} \over {d I_{cm}}} ({I}_{cm} (1, x,1,1)) - 
2 {{d \sigma} \over {d I_{cm}}} ({I}_{cm} (1, 1,1,1)) \Bigg] \nonumber \\
&& + {1 \over 4} 
\beta_f  \Bigg[
{{d \sigma} \over {d I_{cm}}} ({I}_{cm} (1, 1,x,1)) +  
{{d \sigma} \over {d I_{cm}}} ({I}_{cm} (1, 1,1,x)) \nonumber \\
&& -  2 {{d \sigma} \over {d I_{cm}}} ({I}_{cm} (1, 1,1,1)) \Bigg] \Bigg\}  , 
\label{eq:sfca}
\end{eqnarray}
obtained by inserting in eq.~(\ref{eq:sfc}) the distributional expansion of the
structure functions and dropping the spurious higher-order terms. It is worth
noting, for future convenience, that the LL cross sections, both the complete
and the $\cal O (\alpha)$-truncated ones, are functionals of the kernel cross
section. In particular, they define algorithms that, once applied to a generic
kernel, provide the whole ensemble of the LL corrections to all orders 
and the $\cal O (\alpha)$ LL ones, respectively. 
So, when appropriate, we will refer to them as 
$\Sigma_{LL} \Big[ d \sigma / d I  \Big]$ or $\Sigma_{LL}^{(\alpha)} 
\Big[ d \sigma / d I  \Big]$, respectively. 

The matching between the all-orders leading-log cross section,
$\sigma_{LL}$, given by the convolution of structure functions with
kernel (Born) cross sections, and the exact $\cal O(\alpha)$ one
is realized according to
the following general recipe:
the $\cal O (\alpha)$ content of the leading-log
cross section
is extracted by employing the $\cal O(\alpha)$ expansions of the structure
functions, thereby yielding
$\sigma^{(\alpha)}_{LL}$. 
By denoting with $\sigma^{(\alpha)}_{PT}$ the  exact $\cal O (\alpha)$ 
correction to the cross section~\cite{cr, noi}, the fully corrected cross 
section can  be written as
\begin{equation}
\sigma_A  = \sigma_{LL}
 - \sigma^{(\alpha)}_{LL} +
\sigma^{(\alpha)}_{PT}  \; .
\label{eq:add}
\end{equation}

Equation~(\ref{eq:add}), which is in the additive form, is by construction
exact at $\cal O (\alpha)$ and includes higher order corrections as taken into
account by $\sigma_{LL}$. A factorized form can also be
supplied. It has the same $\cal O ( \alpha)$ content but also leads to
the so-called classical limit, according to which the cross section
must vanish in the absence of photonic radiation. It reads
\begin{eqnarray}
&&\sigma_F  = ( 1 + C_{NL}^H ) \sigma_{LL} , \nonumber \\
&&\nonumber \\
&&C_{NL}^H \equiv {{\sigma^{(\alpha)}_{PT}
- \sigma^{(\alpha)}_{LL} }\over \sigma_0} \equiv
{{\sigma_{NL}^{(\alpha)}}\over{\sigma_0}} ,
\label{eq:fact}
\end{eqnarray}
$\sigma_0$ being the Born cross section; $C_{NL}^H$ contains the non-log 
part of
the $\cal O ( \alpha)$ cross section, represented by
$\sigma_{NL}^{(\alpha)}$.

In order to clarify the physical content of eqs.~(\ref{eq:add}) and
(\ref{eq:fact}), it is worth considering the perturbative expansions of their
ingredients, which can be written as follows:  
 
\begin{eqnarray}
&&  \sigma_{LL} = \sigma_0 +  \sigma^{(\alpha)}_{LL} +  
\sigma^{(\alpha^2)}_{LL} +  \sigma^{(\alpha^3)}_{LL} + \ldots ; \nonumber \\
&& \sigma_{LL \vert_{\alpha}} = \sigma_0 +  \sigma^{(\alpha)}_{LL}; \nonumber \\
&& \sigma_{PT} =  \sigma_0 + \sigma^{(\alpha)}_{PT} .
\label{eq:ptexp}
\end{eqnarray}

The dominant part of the difference between the additive cross section and 
the factorized one can be read off the perturbative expansions above: 

\begin{eqnarray}
&&  \sigma_A =  \sigma_0 +  \sigma^{(\alpha)}_{PT} +  
\sigma^{(\alpha^2)}_{LL} +  \sigma^{(\alpha^3)}_{LL} + 
\cal O (\alpha^4 L^4) ; \nonumber \\
&& \sigma_F = \sigma_A + C_{NL}^H \sigma^{(\alpha)}_{LL} 
+ \cal O (\alpha^3 L^2). 
\label{eq:fma}
\end{eqnarray}
The cross-section $\sigma^{(\alpha)}_{LL}$ describes the LL 
universal part of the $\cal O (\alpha)$ corrections,  
and $C_{NL}^H$ the non-log (process dependent) part of the  $\cal O (\alpha)$
 corrections. Therefore, the difference between the factorized 
cross-section $\sigma_F$ and the additive cross-section $\sigma_A$ 
 is due to $\cal O(\alpha^2 L)$ sub-leading corrections 
contained in $\sigma_F$ and not in $\sigma_A$. In particular, the term given by 
$C_{NL}^H \sigma^{(\alpha)}_{LL} $ gives an approximation of the $\cal O
(\alpha^2 L)$ contributions to the cross section, 
\begin{equation}
\sigma^{(\alpha^2 L)} \simeq  C_{NL}^H \sigma^{(\alpha)}_{LL} , 
\label{eq:a2lapp}
\end{equation}
since it is the direct
product of the LL $\cal O (\alpha L)$ cross section times the non-logarithmic
non-universal $\cal O (\alpha)$ correction $C_{NL}^H $. A more detailed
description of the $\cal O (\alpha^2 L)$ correction would require a
convolution of the non-universal non-logarithmic contribution, given by the
difference between the exact $\cal O (\alpha )$ cross section minus the LL 
$\cal O (\alpha )$ one, with 
%initial state 
real+virtual radiation in the LL
approximation. This can be achieved by applying the algorithm of 
eq.~(\ref{eq:sfca}) in the following way: 

\begin{equation}
\sigma^{(\alpha^2 L)} = \Sigma_{LL}^{(\alpha)} \Bigg[
{{d \sigma^{(\alpha)}_{ex}} \over {d \Omega}}
- {{d \sigma^{(\alpha)}_{LL}} \over {d \Omega}} \Bigg]  , 
\label{eq:a2l}
\end{equation}
where $ d \sigma / d \Omega $ are the differential distributions of the
electron scattering angle, already integrated over the photonic phase-space.  
Equations~(\ref{eq:a2lapp}) and~(\ref{eq:a2l}) are, at a first sight, 
very different from one
another. On the contrary, eq.~(\ref{eq:a2l}) reduces to eq.~(\ref{eq:a2lapp})
under the assumption that the ratios between the $\cal O (\alpha )$ densities
and the born one are smooth functions of the cm energy and scattering angle,
which is {\it a priori} a reasonable assumption since the $\cal O (\alpha )$
densities are infrared-dominated and hence almost factorized over the born one.
Anyway, one of the purposes of the present note is just to check the quality of
the approximated description of eq.~(\ref{eq:a2lapp}) and  improve it
by means of eq.~(\ref{eq:a2l}). Some illustrative numerical results will be shown
later on. 

Equation~(\ref{eq:a2l}) performs the $\cal O (\alpha L )$ corrections to the 
$\cal O (\alpha )$ non-logarithmic kernel, and hence provides $\cal O (\alpha^2
L)$ corrections, taking into account also convolution effects. 
Along the same way, one can improve it to
take into account also the $\cal O(\alpha^n L^{n-1})$ corrections in the SF
approach, by computing, instead of~(\ref{eq:a2l}), its improved version, namely

\begin{eqnarray}
&&\sigma^{(\alpha^n L^{n-1})}   = \int d x_1 d x_2 d y_1 d y_2 \int d I_{cm} 
D_{\beta_i}(x_1) D_{\beta_i}(x_2) D_{\beta_f}(y_1) D_{\beta_f}(y_2)
\nonumber \\
&&\Bigg[ \Bigg( 
{{d \sigma^{(\alpha)}_{ex}} \over {d I_{cm}}} ({I}_{cm} (x_1, x_2, y_1 , y_2)) 
- {{d \sigma^{(\alpha)}_{ex}} \over {d I_{cm}}} ({I}_{cm} (1, 1,1,1))  
\Bigg) \nonumber \\
&& -  \Bigg( 
{{d \sigma^{(\alpha)}_{LL}} \over {d I_{cm}}} ({I}_{cm} (x_1, x_2, y_1 , y_2)) 
- {{d \sigma^{(\alpha)}_{LL}} \over {d I_{cm}}} ({I}_{cm} (1, 1,1,1))  
\Bigg) \Bigg] . 
\label{eq:anlnm1}
\end{eqnarray}

%\begin{figure}[hbtp]
%\begin{center}
%\epsfig{file=a2nl.eps,height=11truecm}
%\end{center}
%\label{fig:sabspv}
%\caption{Comparison of old and new versions of the Monte Carlo SABSPV. 
%The relative differences between the different versions 
%involved in the comparison 
% and the cross section by BHLUMI taken as a reference cross section are
%shown as functions of the cut $ z_{min} = E_+ E_- / E_{beam}^2 $. 
%$ E_{+,-} $ are the energies deposited in the positron and electron clusters, 
%respectively. The details of the clustering algorithm (BARE1) are given 
%in~\protect\cite{bharep}. The centre of mass energy is 
%$ \protect\sqrt{s} = 92.3 $~GeV. }
%\end{figure}

Some comments are in order here. If one inserts into eq.~(\ref{eq:anlnm1})
the $\cal O (\alpha^0)$ contribution of the distributional expansion of the
structure function, the result is automatically zero, confirming that 
eq.~(\ref{eq:anlnm1}) contains contributions starting from  
$\cal O (\alpha^2 L)$. 
If one inserts in it 
the $\cal O (\alpha)$ contribution of the distributional expansion of the
structure function, one naturally recovers eq.~(\ref{eq:a2l}). The higher-order
contributions in the structure functions produce  
$\cal O (\alpha^n L^{n-1})$ corrections: they are complete at the LL level
until $\cal O (\alpha^3 L^{2})$, correct in the soft limit for all the
higher-order next-to-leading contributions due to the fact that at present the
structure functions employed take into account only up to 
$\cal O (\alpha^2 )$  hard-photon radiation. This, of course, is not a
limitation in principle, and moreover has no practical influence for the
usually adopted experimental cuts. 

The discussion up to now, together with the definition of the various
ingredients, allows the definition of a new, factorized cross section. In fact,
it is easy to verify that the sum of the additive cross section of 
eq.~(\ref{eq:add}) and the higher-order next-to-leading correction of 
eq.~(\ref{eq:anlnm1}) defines the following factorized cross section: 

\begin{eqnarray}
&&\sigma_A +  \sigma^{(\alpha^n L^{n-1})} = \sigma^{new}_{F} , \nonumber \\
&&\sigma^{new}_{F} = \int d x_1 d x_2 d y_1 d y_2 \int d \Omega_{cm} 
D_{\beta_i}(x_1) D_{\beta_i}(x_2) D_{\beta_f}(y_1) D_{\beta_f}(y_2) 
\nonumber \\
&&\quad {{d \sigma_0} \over {d \Omega_{cm}}} (x_1, x_2, y_1 , y_2)  
 \Bigg( 1 + C_{NL}^H (x_1, x_2, y_1 , y_2), 
 \Bigg) , \nonumber \\
&&C_{NL}^H (x_1, x_2, y_1 , y_2)  = {{
d \sigma^{(\alpha)}_{ex} / d \Omega_{cm} 
- d \sigma^{(\alpha)}_{LL} / d \Omega_{cm}
} \over {d \sigma_0 / d \Omega_{cm}}}  . 
\label{eq:newf}
\end{eqnarray}

The analytical identity between $\sigma_A +  \sigma^{(\alpha^n L^{n-1})}$ on
the one hand, and $\sigma^{new}_{F}$ on the other one can be seen as follows.
$\sigma^{new}_{F}$ is the sum of two integrals. The first one, involving only
the born-approximation density, is nothing but $\sigma_{LL}$. The second one,
involving the product of the  born-approximation density times the new 
$ C_{NL}^H (x_1, x_2, y_1, y_2) $, coincides with 
the full convolution of the difference between the exact $\cal O (\alpha )$
density and the LL-approximated one. Let us focus the attention on this last
integral. By inserting in it the $\cal O (\alpha^0 )$ distributional expansion
of the structure functions and performing the integrations, one is left with
$ \sigma^{(\alpha)}_{PT}  - \sigma^{(\alpha)}_{LL}$ which, when
summed up to $\sigma_{LL}$, gives exactly $\sigma_A$.  The residual
higher-order contributions to the second integral, by virtue of the fact that
the structure functions are normalized to unity, recover exactly 
$\sigma^{(\alpha^n L^{n-1})}$. 

Equation~(\ref{eq:newf}) reproduces the result of eq.~(\ref{eq:fact}) under the
assumption that 
$ C_{NL}^H (x_1, x_2, y_1, y_2) $ is a smooth function
of its arguments, so that it can be extracted from the convolution integral and
the densities appearing in its  numerator and denominator can be replaced by
the corresponding integrated cross sections. In this sense, eq.~(\ref{eq:newf})
is an improvement of eq.~(\ref{eq:fact}), that fully takes into account the
angle and energy dependence of the non-logarithmic component of the exact 
$\cal O (\alpha)$ correction, so that eq.~(\ref{eq:newf})  supersedes 
eq.~(\ref{eq:fact}) developed in ref.~\cite{sabsyr}, 
which the published version of the FORTRAN code SABSPV~\cite{sabscpc} is based
on.   
%A first approach to this problem has been
%done 
Moreover, from a technical point of view, the formulation described 
in~\cite{sabsyr} is worked out under the assumption that a specific 
cut on the final
state fermions is applied, namely a cut on the individual 
energies of the fermions. 
% and
%the results presented in~\cite{bharep} have been obtained by approximating the
%final-state energy cuts in terms of cuts on the individual energies of the
%final-state fermions. 
The present formulation allows one, on the contrary, to impose arbitrary 
cuts on the final state fermions.

The formulation just described is conceived for applications to the SABH
process. Nonetheless, it is completely general, and can be applied also to
annihilation processes, once the non-logarithmic $\cal O (\alpha)$
process-dependent correction ($C_{NL}^H$) is known. 

It is worth mentioning that another source of $\cal O (\alpha^2 L )$ 
corrections is represented by the production of additional light pairs; 
this contribution is at present well under control at the level of 
0.03\%~\cite{bharep,comdoc}, 
and a further improvement will be necessary only when 
photonic corrections will be under control at the same level. 

So far, only QED corrections to the dominant part of the
small-angle Bhabha cross section, i.e. the $t$-channel photon exchange
contribution,  have been considered. This is not a limitation of the approach:
the very same algorithm of eq.~(\ref{eq:newf}) can be extended to the full
Bhabha cross section in a straightforward way. Actually, from the practical
point of view, at small scattering angle (20--60~mrad) it is necessary to take
into account the residual born-approximation contributions plus their 
LL corrections, and this can be simply obtained by adding the residual
born-approximation contributions to the integration kernel of 
eq.~(\ref{eq:newf}). Moreover, the only non-QED corrections relevant in the
luminometry region are given by the photonic vacuum polarization, which also
can be easily taken into account by using the running QED coupling constant. 

\vskip 12pt\noindent
\leftline{\large\bf 3. Comparisons and numerical results}
%\vskip 12pt
%
%\leftline{\it -- Analytical comparisons}
%\vskip 8pt

%\noindent
{\it Analytical comparisons} -- The approach just described 
insures that the $\cal O (\alpha^2 L )$ photonic 
contributions coming from an ``external'' collinear photon in association with
an ``internal'' non-collinear one are automatically
taken into account. This is not in principle 
the complete set of $\cal O (\alpha^2 L )$ corrections: in this way, for
instance,  the truly
irreducible two-loop corrections are missed, but they can be expected to give
small contributions. 
%Actually, one could also have an ``internal'' collinear photon
%accompanied by an ``external'' non-collinear one. But, in this case, in order
%to have an $\cal O (\alpha^2 L )$ contribution,  the ``external'' 
%photon is forced to be confined to the soft region of the phase space, because 
%otherwise the internal fermionic leg is off-shell and hence cannot provide 
%a collinear logarithm. 
Therefore, the $\cal O (\alpha^2 L )$ corrections
taken into account by this method represent the bulk of the complete set. This
heuristic argument can be put on a firmer ground by comparing the results
provided by this formulation  with the ones obtained by means of complete 
 $\cal O (\alpha^2 L )$ calculations, already present in the literature for
 some specific ES's. 
As a first step, we will compare our results  with the ones available in
analytical form, i.e. in the soft approximation limit; next we will consider
comparisons beyond the soft approximation, for which only numerical results can
be compared.  

A first analytical result including $\cal O (\alpha^2 L )$ corrections concerns
$e^+ e^- $ annihilation into $\mu^+ \mu^- $ pairs, taking into account 
initial-state QED corrections~\cite{kf}. In that paper an analytical formula is
given, 
describing QED corrections for an ES  where  the total energy emitted by
initial-state photons does not exceed  $\Delta E$. If one works out
eq.~(\ref{eq:newf}), namely by putting in this case $\beta_f = 0$, 
imposing the cut condition of~\cite{kf}, computing the proper $C_{NL}^H$ 
and truncating the result at the 
$\cal O (\alpha^2)$,  one finds that:  
\begin{itemize}
\item the $\cal O (\alpha )$ perturbative result is exactly recovered, by
construction; 
\item \underline{all} the infrared-singular terms,
namely the ones containing $\ln^2  \varepsilon  $ and 
$\ln  \varepsilon  $, where $\varepsilon = \Delta E / E$,  
are exactly recovered at the level of $\cal O
(\alpha^2 L_s^2)$,  $\cal O (\alpha^2 L_s)$ and $\cal O (\alpha^2 )$, 
where $L_s = \ln (s / m^2 )$; 
\item the difference between the two
results starts at the level of $ (\alpha / \pi )^2 L_s $ times a constant;    
\end{itemize} 
in particular, such a difference reads
\begin{equation}
{{\delta \sigma} \over {\sigma_0}} \Bigg\vert_{(\alpha^2 L_s)}= 
\left( {\alpha \over \pi} \right)^2 L_s \left[ 
3 \zeta (3) - {3 \over 2} \zeta (2) + {3 \over 16} \right],
\label{eq:schdiff} 
\end{equation}
where $\delta \sigma $ is the difference between the cross section
of~\cite{kf} and the cross section of eq.~(\ref{eq:newf}). 
The difference  numerically amounts to a relative deviation of about 
 $1.7 \times 10^{-4}$. The residual difference is 
 at $\cal O (\alpha^2) $ times a constant and is numerically irrelevant. 
 
It is worth noting that results including full $\cal O (\alpha^2 L )$
corrections for the small-angle Bhabha scattering cross section are available 
for an academic trigger~\cite{nllbha} or under development for the most general
case~\cite{a2lbrem}. Also in this case it is possible to compare analytically
the predictions of eq.~(\ref{eq:newf}) with the results shown in~\cite{nllbha}
in the soft approximation. The comparison can be worked out along the same
lines as in the previous case, with the only differences that the collinear
logarithm is now $L_t = \ln (-t / m^2)$, $\beta_f = \beta_i = \beta_t$ and the
cut condition imposed in~\cite{nllbha} requires that the energy of each photon
does not exceed $\Delta E$. The results of the comparison are the same as
before up to the $\cal O (\alpha^2 L_t )$ corrections,  
namely the difference appears at the level of $ (\alpha / \pi )^2 L_t $ 
times a constant and reads
\begin{equation}
{{\delta \sigma} \over {\sigma_0}} \Bigg \vert_{(\alpha^2 L_t)}= 
2 \left( {\alpha \over \pi} \right)^2 L_t \left[ 
3 \zeta (3) - {3 \over 2} \zeta (2) + {3 \over 16} \right],
\label{eq:tchdiff} 
\end{equation}
where $\delta \sigma $ is the difference between the cross section
of~\cite{nllbha} in soft approximation 
and the cross section of eq.~(\ref{eq:newf}).
This difference numerically amounts  to a relative deviation of about 
$2.2 \times 10^{-4}$, since the overall factor of two is compensated by the
fact that $L_t \simeq 2/3 L_s$. In this case, also an additional difference
appears, namely at the level of the infrared-sensitive truly 
$\cal O (\alpha^2)$ terms, which reads
\begin{equation}
{{\delta \sigma} \over {\sigma_0}} \Bigg \vert_{((\alpha^2 ))}= 
- \left( {\alpha \over \pi} \right)^2  \left[ 4 \ln^2 \varepsilon + 8 \ln
\varepsilon \right], 
\label{eq:irdiff}
\end{equation} 
the residual difference being at $\cal O (\alpha^2) $ times a constant and 
hence numerically irrelevant. This last difference, being infrared-sensitive, 
can show up in the region of strong cuts.

%It is worth mentioning that another source of $\cal O (\alpha^2 L )$ 
%corrections is represented by the production of additional light pairs; 
%this contribution is at present well under control at the level of 
%0.03\%~\cite{bharep}, and a further improvement will be necessary only when 
%photonic corrections will be under control at the same level. 

%So far, only QED corrections to the dominant part of the
%small-angle Bhabha cross section, i.e. the $t$-channel photon exchange
%contribution,  have been considered. This is not a limitation of the approach:
%the very same algorithm of eq.~(\ref{eq:newf}) can be extended to the full
%Bhabha cross section in a straightforward way. Actually, from the practical
%point of view, at small scattering angle (20--60~mrad) it is necessary to take
%into account the residual born-approximation contributions plus their 
%LL corrections, and this can be simply obtained by adding the residual
%born-approximation contributions to the integration kernel of 
%eq.~(\ref{eq:newf}). Moreover, the only non-QED corrections relevant in the
%luminometry region are given by the photonic vacuum polarization, which also
%can be easily taken into account by using the running QED coupling constant. 
%\vskip 12pt
%\leftline{\it -- Numerical results}
%\vskip 8pt 

%\noindent
{\it Numerical results} -- Going beyond the soft 
approximation requires the discussion of numerical
results.  From now on, it
is understood that all the numerical results, apart from the ones obtained by
our new formulation, can be found in~\cite{bharep}.  
The formulation described in this paper has been implemented in a new,
so far unpublished, version of the code SABSPV~\cite{sabscpc}. 

%%%%%%%%%%%%%%%%%%%%%%%BARE1%%%%%%%%%%%%%%%%%%%%%%%%%%%%%%%%%%%%%%

\begin{figure}[hbtp]
\begin{center}
\epsfig{file=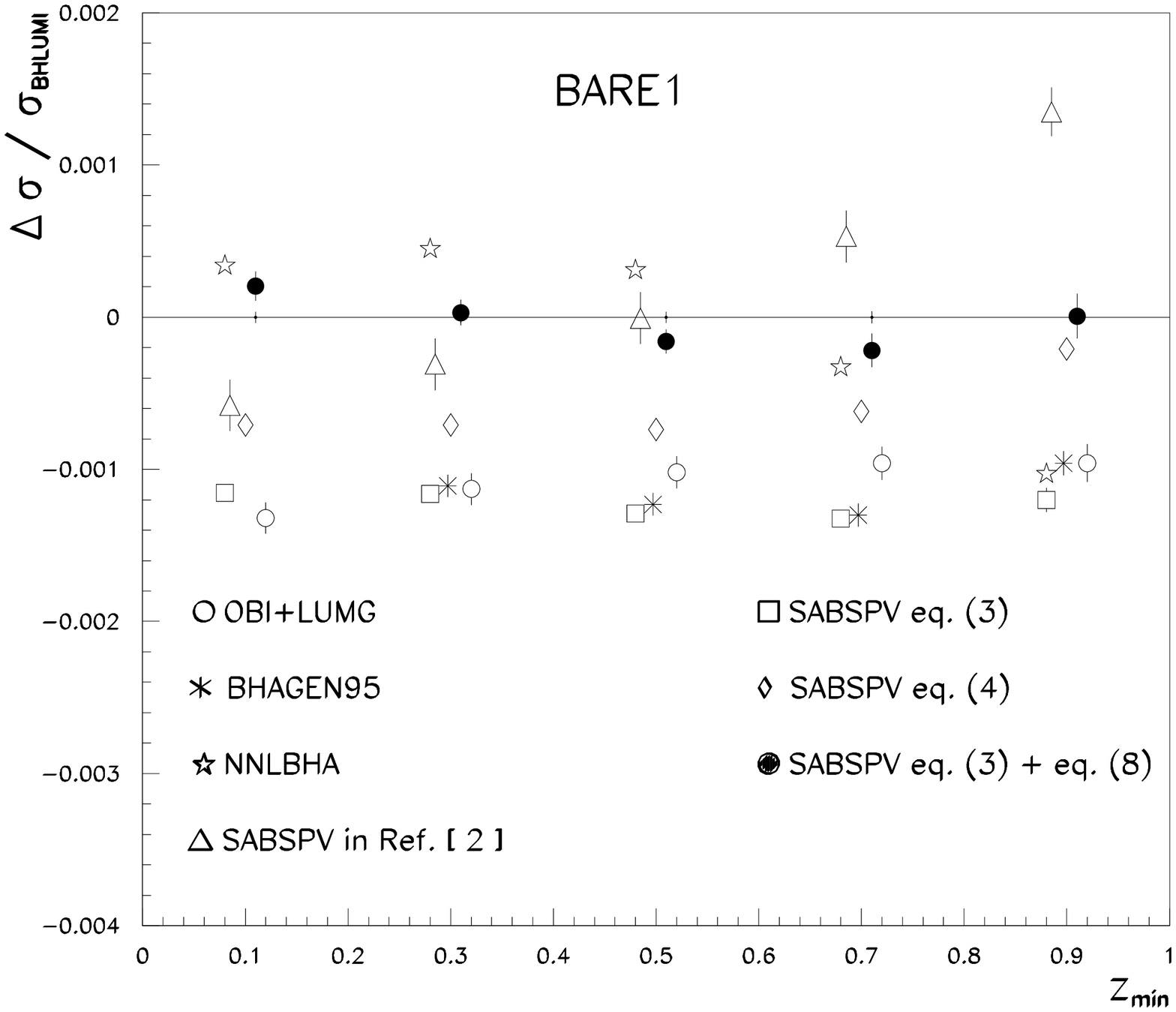,height=11truecm}
\end{center}
\caption{Comparison of MC and non-MC results. 
The relative differences between the codes 
involved in the comparison 
and the cross section by BHLUMI taken as a reference cross section are
shown as functions of the cut $ z_{min} = E_+ E_- / E_{beam}^2 $. 
$ E_{+,-} $ are the energies deposited by the bare final state 
positron and electron, 
respectively. The details of the clustering algorithm (BARE1) are given 
in~\protect\cite{bharep}. The centre of mass energy is 
$ \protect\sqrt{s} = 92.3 $~GeV. }
\label{fig:bare1}
\end{figure}

Figure~\ref{fig:bare1} shows the comparison between 
the results obtained by the
present formulation of the problem and the results available in the
literature~\cite{bharep} for a Bare ES. For
the details concerning the ES, the reader is referred to~\cite{bharep}. The
FORTRAN codes involved in the comparison are BHAGEN95~\cite{bhagen},
BHLUMI~\cite{bhlumi}, NLLBHA~\cite{nllbha}, 
OLDBIS+LUMLOG~\cite{bhlumi} and 
SABSPV~\cite{sabsyr,sabscpc}. 
Again, the details concerning the codes and their
underlying theoretical formulations can be better found in the original
literature and in~\cite{bharep}. Figure~\ref{fig:bare1} is drawn according to
the convention adopted in~\cite{bharep}, namely it shows the relative
differences between the codes involved and a reference cross section, which in
the present case is the cross section computed by BHLUMI. 
As a first comment, one can see that 
our new factorized solution corresponding to eq.~(\ref{eq:fact}) differs 
appreciably from the previous factorized one of ref.~\cite{sabsyr} due 
to the improved treatment of the final state radiation phase space.
Both of these solutions, neglecting convolution effects 
at the level of $\cal O (\alpha^2 L)$ corrections, are at present obsolete.
By looking at the figure, one
can see that all the solutions shown group together essentially into two
clusters. The first cluster of solutions is the one of the ``additive''
solutions, namely BHAGEN95, OLDBIS+LUMLOG and SABSPV in its additive version
(see eq.~(\ref{eq:add})). All these solutions have the common feature that they 
miss the $\cal O (\alpha^2 L)$ contributions, since they do not fill the region
of phase space characterized by the emission of one collinear and one
acollinear photon. Their relative distance from the reference cross section 
is steadily around 0.10--0.15\%. 
%growing to around 0.5\% (and beyond) in the region of strong cuts.
The second cluster of solutions is the one of the ``factorized''
solutions, namely BHLUMI and SABSPV in its factorized versions (see
eqs.~(\ref{eq:fact}) and (\ref{eq:newf})).\footnote{\footnotesize Actually,
 the results shown in Fig.~\ref{fig:bare1} have been obtained by means of the
sum of eq.~(\ref{eq:add}) and eq.~(\ref{eq:a2l}), which anyway differs from
eq.~(\ref{eq:newf}) starting from $\cal O (\alpha^3 L^2)$ corrections; these
corrections are numerically irrelevant, as checked for the case under
consideration, and shown later on for a realistic ES.} 
All these solutions have the common 
feature that they include the bulk of the $\cal O (\alpha^2 L)$ contributions
in some form. It is worth noting that all of them lye in the band
$\sigma_{ref} \pm 0.1\%$.  The comparison between the results of
eqs.~(\ref{eq:fact}) and (\ref{eq:newf}) is a measure of the degree of
approximation of eq.~(\ref{eq:fact}) with respect to the new improved solution
of eq.~(\ref{eq:newf}). The relative difference between them is maximum at 
$ z_{min} = 0.1$, where it is roughly 0.1\% and can be attributed to
convolution effects, and minimum at  $ z_{min} =
0.9$, where it amounts to around 0.02\%  since the convolution effects 
switch off naturally.  
These differences should  be compared with the
corresponding relative difference with the cluster of additive solutions, 
which on the contrary is larger than 0.1\%. 
This means that the approximate solution of 
eq.~(\ref{eq:fact}) is, after all, a good approximation to eq.~(\ref{eq:newf}),
which anyway supersedes it. It is worth stressing that the main difference
between eq.~(\ref{eq:fact}) and (\ref{eq:newf}) does not lye in the dynamical
content, but rather in convolution and/or phase-space effects, that are
correctly treated in (\ref{eq:newf}) while approximated in (\ref{eq:fact}). 
Note that the new improved solution and the
reference cross section differ from one another by less than 0.03\% over all
the  $z_{min}$ range. 
As far as the physical content is concerned, the results by NLLBHA are
comparable with the factorized results by BHLUMI and SABSPV. In particular,
also these results differ from the reference cross section by less than 0.1\%
over all the $z_{min}$ range and less than 0.06\% for realistic  $z_{min}$,
namely $0.3 \leq z_{min} \leq 0.7$. The maximum difference between NLLBHA and
BHLUMI/SABSPV is at $z_{min} = 0.9$, and could be traced back  to the 
infrared-sensitive terms of eq.~(\ref{eq:irdiff}). This item deserves further
investigation.

%As in the case of the Calorimetric ES, 
%the cluster of the additive solutions differ from the reference cross section 
%by about 0.1\%, almost independently of the $z_min$ cut. Also a cluster of 
% solutions can be seen, namely  
%the results corresponding to eq.~(\ref{eq:newf}), plus the results of NNLBHA,
%which differ by less than 0.05\% from the reference
%cross section for realistic $z_{min}$ cuts. 

%%%%%%%CALO2%%%%%%%%%%%%%%%%%%%%%%%%%%%%%%%%%%%%%%%%%
%????????????????????????????

\begin{figure}[hbtp]
\begin{center}
\epsfig{file=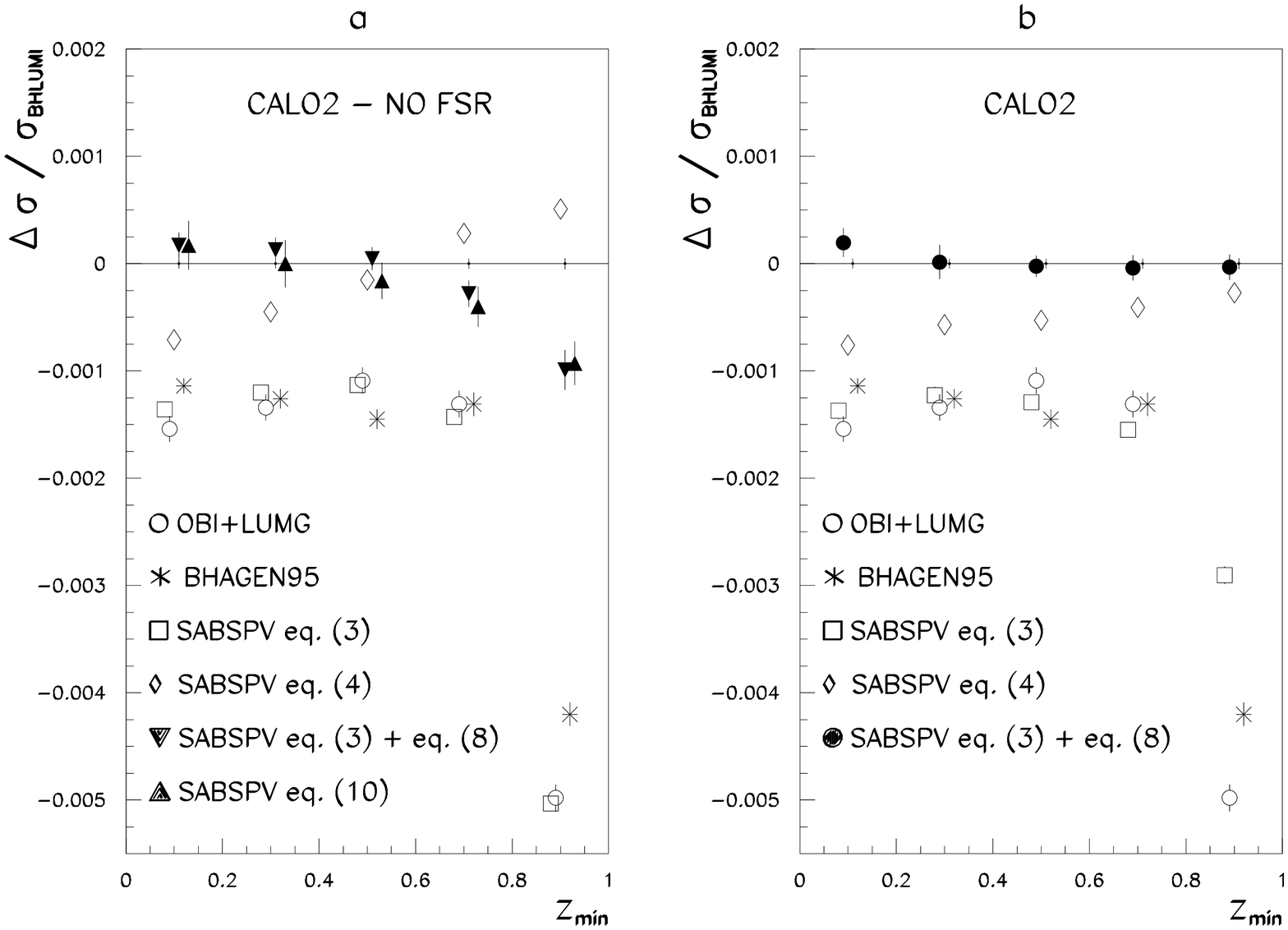,height=11truecm}
\end{center}
\caption{Comparison of Monte Carlo's. 
The relative differences between the codes 
involved in the comparison 
and the cross section by BHLUMI taken as a reference cross section are
shown as functions of the cut $ z_{min} = E_+ E_- / E_{beam}^2 $. 
$ E_{+,-} $ are the energies deposited in the positron and electron clusters, 
respectively. The details of the clustering algorithm (CALO2) are given 
in~\protect\cite{bharep}. The centre of mass energy is 
$ \protect\sqrt{s} = 92.3 $~GeV. }
\label{fig:calo2}
\end{figure}

The formulation described above in eq.~(\ref{eq:newf}) is conceived for a
Bare ES. Going to a Calorimetric ES requires a completely different treatment
of final-state radiation in the LL part of the result. A first step can be 
simply dropping final-state radiation, which means putting $\beta_f = 0 $ in
all the structure function corrections, consistently with the fact that,  
for an inclusive  measurement, final-state
mass ``singularities'' are absent.  An improvement with respect to this choice
can be taking into account final-state radiation having defined a proper
final-state radiation factor. This can be done by considering that adding the
correction due to photons collinear to final-state leptons results into a
modification of the $\beta $ factor~\cite{coll} according to 
\begin{equation}
\beta_t = 2 (\alpha / \pi) \left[ \ln (-t / m^2 ) - 1 \right] \to 
\beta_\delta = 2 (\alpha / \pi)  
\ln \left( {{- 4 t} \over  {\delta^2 s}}  \right) , 
\label{eq:delc} 
\end{equation}
where $\delta$ can be reasonably taken as the minimum aperture of the
final-state cluster.  
The results obtained for a CALO2 ES (see \cite{bharep} for details) are shown in
Fig.~\ref{fig:calo2}, where the same conventions as in Fig.~\ref{fig:bare1}
have been adopted. The only difference is that, in this case, the results by
NLLBHA are absent and the numerical results relative to the 
present formulation have been produced first switching off final-state radiation 
(Fig.~2a),  and then including its effect using in the structure 
functions $\beta_f = \beta_\delta$ with $\beta_\delta$ 
given by~(\ref{eq:delc}) (Fig.~2b). 
As in the case of the BARE ES, two clusters of solutions can be recognized.  
The first group of the additive solutions, due to the lack of $\cal O (\alpha^2 L)$ 
 contributions, differ from the reference cross section of 
 about 0.10-0.15\% for $0.1 \leq z_{min} \leq 0.7$ and of about 0.3-0.5\%  
 at the extreme value $z_{min} = 0.9$, where the differences can be expected to be 
 enhanced by infrared-sensitive terms. The second cluster of the
 ``factorized'' 
 solutions (BHLUMI and SABSPV in its factorized versions), which include 
 the bulk of the $\cal O(\alpha^2 L)$ corrections,
  is contained within a band $\sigma_{ref} \pm 0.1\%$, 
 both without and with final-state radiation. In particular, 
 for the case of no final-state radiation (Fig.~2a), we show also 
 the numerical results 
 corresponding to the sum of eq.~(\ref{eq:add}) with eq.~(\ref{eq:a2l}), which, 
 although in additive form, does contain $\cal O (\alpha^2 L)$ contributions. 
 As can be seen, the difference between the new factorized 
 version of SABSPV of eq.~(\ref{eq:newf}) and the result given by the 
 sum eq.~(\ref{eq:add}) with  eq.~(\ref{eq:a2l}) is not appreciable, even at 
 this 
 level of precision, since the two prescriptions differ for sub-leading
  terms starting from $\cal O (\alpha^3 L^2)$. 
  The largest difference between 
  the improved versions of SABSPV and the reference cross section is present 
  at $z_{min} = 0.9$ in Fig.~2a, 
  and it has to be ascribed to neglecting final-state 
  radiation corrections in SABSPV, for this case. 
  This interpretation is confirmed by 
 the  situation shown in Fig.~2b,
  where the predictions by SABSPV include 
 the effect of final-state radiation. Actually, the new improved solution and 
 the reference cross section differ, in this case, by 
 less than 0.02\% over the full $z_{min}$
  range, similarly to the case of the BARE ES. 
  As a general comment, it is worth 
  noting that, wherever differences between  solutions 
  are present at the extreme value $z_{min} = 0.9$, these are due, as we 
  explicitly shown independently of the ES, to differently treated 
  infrared-sensitive contributions.

%Figure~(\ref{fig:bare1})
%It is not a limitation in principle: the formulation
%can be extended to describe also Bare ES's, provided one includes also the
%effect of final state radiation in the LL part. Actually this extension has
%been done along the same lines as the calorimetric
%formulation by following a lengthy but straightforward procedure which will be
%shown elsewhere~\cite{npbbha}.

%??????????????????????????????????

In conclusion, we have shown a new formulation for the  calculation
of photonic corrections to a generic kernel cross section. 
A method for computing the bulk of photonic $\cal O (\alpha^2 L)$
corrections has been proposed, and analytically tested versus theoretical
results present in the literature for simple ES's. The algorithm has been
applied to the precision calculation of the SABH cross section, 
relevant for the luminosity measurement at $e^+ e^-$
colliders, and several numerical results have been 
shown and critically commented. 

The present formulation, as applied to the SABH process, 
supersedes the one developed 
in~\cite{sabsyr, sabscpc} and used in~\cite{bharep,comdoc} 
as one of the tools for
the estimate of the  theoretical error on the SABH process itself. 

Following the strategy adopted in~\cite{bharep,comdoc}, 
an appreciable reduction  of such an error will only be
reached by means of a critical comparison of all the available formulations of
the problem, for a wide ensemble of ES's.  
The present formulation  is able to provide one of the 
ingredients for such an achievement, towards a theoretical error on 
luminosity well below 0.1\%.    

\vskip 24pt \noindent
{\bf Acknowledgements} The authors are indebted to S.~Jadach and B.F.L.~Ward 
for stimulating discussions and comments on the subject. Special thanks are
due to S.~Jadach for having encouraged the present study. 

%%%%%%%%%%%%%%%%%%%%%%%%%%%%%%%%%%%%%%%%%%%%%%%%%%%%%%%%%%%%%%%%%%%%%%%%% 

%\vfill\eject
%\leftline{\large \bf Figure Captions}
%\vskip 30pt
%\noindent
%Figure 1. 

\begin{thebibliography}{9}

\bibitem{experr} B.~Pietrzyk, in {\it Tennessee International Symposium on
Radiative Corrections: Status and Outlook}, edited by B.F.L.~Ward (World
Scientific, Singapore, 1995), Gatlinburg, Tennessee, USA, June 1994; \\
LEP Electroweak Working Group, ``A Combination of Preliminary LEP Electroweak
Results from the 1995 Summer Conferences'', 1995, CERN Report LEPEWWG/95-02; \\
LEP Collaborations, 1995, Collaboration Notes: ALEPH 95-093 PHYSICS 95-086;
DELPHI 95-137 PHYS 562; L3 Note 1814; OPAL Technical Note TN312, 1 August 1995.

\bibitem{bharep} {S. Jadach (convener), O.~Nicrosini (convener), 
H.~Anlauf, A.~Arbuzov, M.~Bigi, H.~Burkhardt, M.~Cacciari, M.~Caffo, H.~Czy\.z,
M.~Dallavalle, J.~Field, F.~Filthaut, F.~Jegerlehner, E.~Kuraev, G.~Montagna,
T.~Ohl, F.~Piccinini, B.~Pietrzyk, W.~P\l{}aczek, E.~Remiddi, M.~Skrzypek,
L.~Trentadue, B.~F.~L.~Ward, Z.~W\c{a}s, ``Event Generators for Bhabha 
Scattering'', in {\it Physics at LEP2}, 
G.~Altarelli, T.~Sj\"ostrand and F.~Zwirner, eds., 
CERN Report {\bf 96-01} (Geneva, 1996), vol.~2, p.~229.  }

\bibitem{comdoc} {A.~Arbuzov, M.~Bigi, H.~Burkhardt, M.~Cacciari, 
M.~Caffo, H.~Czy\.z,
M.~Dallavalle, J.~Field, F.~Filthaut, S.~Jadach, F.~Jegerlehner, E.~Kuraev, 
G.~Montagna, O.~Nicrosini, F.~Piccinini, 
B.~Pietrzyk,  W.~P\l{}aczek, E.~Remiddi, M.~Skrzypek,
L.~Trentadue, B.~F.~L.~Ward, Z.~W\c{a}s, ``The present theoretical error on the
Bhabha scattering cross section in the luminometry region at LEP'', 
hep-ph/9605239. }

\bibitem{bhagen} {M.~Caffo, H.~Czy{\.z} and E.~Remiddi,
in {\it Reports of the Working Group on Precision Calculations
for the $Z$ Resonance}, D.~Bardin, W.~Hollik and G.~Passarino, eds., 
CERN  Report {\bf 95-03}
(Geneva, 1995), p.~361; long write-up in~\cite{bharep}. }

\bibitem{bhlumi} {S.~Jadach, E.~Richter-W\c{a}s, B.~F.~L.~Ward and 
Z.~W\c{a}s, Phys.~Lett. {\bf B353} (1995) 362; Comput. Phys. 
Commun. {\bf 70} (1992) 305. }

\bibitem{nllbha} A.~Arbuzov, V.S.~Fadin, E.A.~Kuraev, L.N.~Lipatov,
N.P.~Merenkov and L.~Trentadue, 
``Small angle electron-positron scattering with a per mille accuracy'',  
CERN-TH/95-313, hep-ph/9512344, and
references therein. 

%%%\bibitem{obslmg} {See ref.~\cite{bhlumi}. }

\bibitem{sabsyr} M.~Cacciari, G.~Montagna, O.~Nicrosini and F.~Piccinini,
in {\it Reports of the Working Group on Precision Calculations 
for the $Z$ Resonance}, D.~Bardin, W.~Hollik and G.~Passarino, eds., 
CERN  Report {\bf 95-03} 
(Geneva, 1995), p.~389, and references therein.

\bibitem{sabscpc} M.~Cacciari, G.~Montagna, O.~Nicrosini and F.~Piccinini,
Comput. Phys. Commun. {\bf 90} (1995) 301.

\bibitem{sf} {E.A.~Kuraev and V.S.~Fadin, Sov. J. Nucl. Phys. {\bf 41} (1985)
466; \\
G.~Altarelli and G.~Martinelli, in {\it Physics at LEP}, J.~Ellis and
R.~Peccei, eds., CERN Report {\bf 86-02} (Geneva, 1986), vol.~1, p.~47; \\
O.~Nicrosini and L.~Trentadue, Phys. Lett. {\bf B196} (1987) 551; Z. Phys. {\bf
C39} (1988) 479. }

\bibitem{cr} {M.~Caffo, E.~Remiddi et al., ``Bhabha Scattering'', in {\it 
Z Physics at LEP1}, G.~Altarelli, R.~Kleiss and
C.~Verzegnassi, eds., 
CERN Report {\bf 89--08} (Geneva, 1989),  vol.~1, p.~171, 
and references therein. }

\bibitem{noi}{M.~Greco, G.~Montagna, O.~Nicrosini and F.~Piccinini,
Phys.~Lett.~{\bf B318} (1993) 635 and references therein; \\
G.~Montagna, O.~Nicrosini and F.~Piccinini,
 Comput.~Phys.~Commun. {\bf 78} (1993) 155; erratum, ibid.~{\bf 79}
(1994) 351. }

\bibitem{kf} {See E.A.~Kuraev and V.S.~Fadin in~\cite{sf}. }

\bibitem{a2lbrem} S.~Jadach, M.~Melles, B.F.L.~Ward, S.A.~Yost, ``Exact 
$ O (\alpha) $ Corrections to the Single Hard Bremsstrahlung Process in 
Low Angle Bhabha Scattering in the SLC/LEP Energy Regime'',  UTHEP-95-0101, 
hep-ph/9603248.  

%%%\bibitem{npbbha} G.~Montagna, O.~Nicrosini and F.~Piccinini, in preparation. 

\bibitem{coll} M.~Cacciari, G.~Montagna and O.~Nicrosini, Phys.~Lett. {\bf 
B274} (1992) 473,  and references therein.

\end{thebibliography}
\end{document}